\documentclass[letterpaper]{article} 
\usepackage{aaai2026}  
\usepackage{times}  
\usepackage{helvet}  
\usepackage{courier}  
\usepackage[hyphens]{url}  
\usepackage{graphicx} 
\usepackage{natbib}  
\usepackage{caption} 
\usepackage{algorithm}
\usepackage{algorithmic}

\usepackage{newfloat}
\usepackage{listings}
\DeclareCaptionStyle{ruled}{labelfont=normalfont,labelsep=colon,strut=off} 
\floatstyle{ruled}
\newfloat{listing}{tb}{lst}{}
\floatname{listing}{Listing}
\usepackage{amssymb}
\usepackage{amsmath}
\usepackage{amsthm}
\usepackage{mathtools}
\usepackage{amsfonts}
\usepackage{multirow}
\usepackage{dcolumn}
\usepackage{siunitx}
\usepackage{bm}
\usepackage{array}
\usepackage{longtable}
\usepackage{mathrsfs}
\usepackage{ascmac}
\usepackage{comment}
\usepackage{breqn}
\usepackage{diagbox}
\usepackage{autobreak}
\usepackage{subfig}

\newtheorem{theorem}{Theorem}

\title{
Causal Inference Under Threshold Manipulation: Bayesian Mixture Modeling and Heterogeneous Treatment Effects
}
\author{
    Kohsuke Kubota\textsuperscript{\rm 1},
    Shonosuke Sugasawa\textsuperscript{\rm 2}
}
\affiliations{
    \textsuperscript{\rm 1}NTT DOCOMO, INC., Tokyo, Japan\\
    \textsuperscript{\rm 2}Faculty of Economics, Keio University, Tokyo, Japan \\

    kousuke.kubota.xt@nttdocomo.com, sugasawa@econ.keio.ac.jp\\
}

\begin{document}

\maketitle

\begin{abstract}
Many marketing applications, including credit card incentive programs, offer rewards to customers who exceed specific spending thresholds to encourage increased consumption. 
Quantifying the causal effect of these thresholds on customers is crucial for effective marketing strategy design. 
Although regression discontinuity design is a standard method for such causal inference tasks, its assumptions can be violated when customers, aware of the thresholds, strategically manipulate their spending to qualify for the rewards.
To address this issue, we propose a novel framework for estimating the causal effect under threshold manipulation.
The main idea is to model the observed spending distribution as a mixture of two distributions: one representing customers strategically affected by the threshold, and the other representing those unaffected. 
To fit the mixture model, we adopt a two-step Bayesian approach consisting of modeling non-bunching customers and fitting a mixture model to a sample around the threshold. 
We show posterior contraction of the resulting posterior distribution of the causal effect under large samples.  
Furthermore, we extend this framework to a hierarchical Bayesian setting to estimate heterogeneous causal effects across customer subgroups, allowing for stable inference even with small subgroup sample sizes. 
We demonstrate the effectiveness of our proposed methods through simulation studies and illustrate their practical implications using a real-world marketing dataset.
\end{abstract}


\section{Introduction}
\label{sec:introduction}
Estimating causal effects of thresholds is important for the effective design of marketing strategies.
Many marketing applications, including loyalty programs~\citep[e.g.][]{kivetz2006goal,nastasoiu2021separating} and credit card incentive programs, offer rewards to customers who exceed specific spending thresholds to encourage increased consumption.
Regression discontinuity design~(RDD) is a representative method to estimate the causal effect of thresholds on customer behavior~\citep{hahn2001identification, thistlethwaite1960regression}.
This approach estimates the causal effect by using local randomization around the threshold and is widely applied as one of the most credible research designs for causal inference, particularly because its identification assumptions are considered weak and plausible~\citep{cattaneo2024practical, mccrary2008manipulation}.

However, the local randomization assumption of RDD can be violated when customers, aware of the thresholds, strategically manipulate their spending to qualify for the rewards.
This strategic behavior of customers at the thresholds breaks the local randomization assumption supported by the \textit{continuity} condition~\citep{hahn2001identification},  which states that the potential outcomes of customers just below and above the threshold are the same.
This condition cannot be satisfied when customers can manipulate their behavior to reach the thresholds because they may have different potential outcomes depending on whether they are above or below the threshold~\citep{ishihara2024manipulation, mccrary2008manipulation, lee2008randomized}.
Although the economics literature proposes the bunching estimation~\citep[e.g.][]{bertanha2024bunching} to address this issue, this framework relies on the assumption that a segment of the distribution just before the threshold, where the probability density function becomes zero, is what is transformed into the observed ``bunch'' at the threshold.
Consequently, the practical applicability of these methods can be limited in scenarios where such a zero-density region cannot be assumed, which often arises in marketing applications.

\begin{figure}
\centering
\includegraphics[width=\columnwidth]{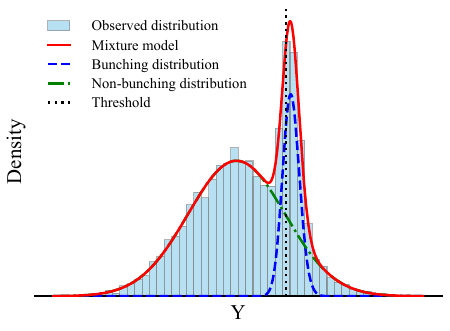}
\caption{Conceptual illustration of our proposed mixture model~(red). Modeling the observed distribution~(skyblue) as a mixture of a non-bunching distribution~(green, unaffected by the threshold) and a strategic bunching distribution~(blue, distorted near the threshold).}
\label{fig:teaser}
\end{figure}

To address causal inference under manipulation, we propose a novel Bayesian framework called Bayesian Modeling of Threshold Manipulation via Mixtures~(BMTM). 
This framework employs a Bayesian mixture model to distinguish between customers strategically affected by a threshold and those who are not, enabling robust causal effect estimation with uncertainty quantification~(see Figure~\ref{fig:teaser}).
To fit the mixture model, we adopt a two-step Bayesian approach consisting of modeling non-bunching customers and fitting a mixture model to a sample around the threshold.
In addition, we extend BMTM to a hierarchical Bayesian setting to estimate heterogeneous causal effects across subgroups of customers, which we call hierarchical BMTM~(HBMTM).
This hierarchical structure ensures stable estimation, even for small subgroups, by borrowing information across them to meet the increasing demand in marketing for estimating heterogeneous causal effects.

We demonstrate the effectiveness of our proposed framework through simulation studies and analysis of a real-world marketing dataset.
The results of the simulation studies indicate that our proposed methods achieve more accurate and reliable estimates of causal effects than conventional RDD methods.
In particular, HBMTM provides stable estimates of heterogeneous causal effects across customer subgroups, even in scenarios with small sample sizes.
The real-world case study further illustrates the practical implications and utility of our framework.

\section{Related Work}
This section reviews related work on bunching estimation and methods to estimate heterogeneous causal effects using RDD.
The bunching estimation examines the discontinuity in the distribution of the running variable.
In the literature, there are two types of discontinuities: kinks, where the slope of an incentive schedule changes~\citep[e.g.][]{saez2010do}, and notches, where the level of the incentive changes discontinuously~\citep[e.g.][]{slemrod2013buenas}.
Marketing incentive programs, which offer a discrete reward upon reaching a threshold, typically fall into the notch category.
In addition to the literature in Section~1, the existing bunching estimation methods~\citep[e.g.][]{kleven2013using} assume sharp bunching, where individuals can precisely target and locate themselves at the threshold.
However, this assumption is often too strong in real-world marketing scenarios because customers may not be fully aware of the exact threshold value or face constraints and imperfect control over their spending behavior.
In contrast to these existing approaches that assume sharp bunching, our proposed framework explicitly models the entire distribution of customer behavior around the threshold.
By employing a mixture model for density estimation, our proposed method naturally captures the diffuse nature of bunching that arises from customers' imperfect control, allowing for causal effect estimation under more plausible assumptions for marketing applications.

In RDD~(without threshold manipulation), there are some methods to estimate the heterogeneous treatment effect across subgroups. 
For example, \citet{sugasawa2023hierarchical} developed a hierarchical model to estimate the subgroup-specific causal effect, and \citet{alcantara2024modified} and \citet{tao2025bayesian} employed the Bayesian additive regression tree to capture heterogeneous effects. 
However, these methods cannot be applied under threshold manipulation. 
To our knowledge, no attempts have been made to estimate the heterogeneous causal effects under threshold manipulation.
Therefore, this paper would be the first to provide an effective method for estimating heterogeneous causal effects under threshold manipulation.

\section{Bayesian Modeling of Threshold Manipulation via Mixtures}

\subsection{Causal Inference Under Threshold Manipulation}
Consider the problem of estimating the causal effect of a threshold $K$ on customers with the potential outcome framework~\citep{imbens2015causal,rubin1974estimating}.
Let $Y(1)$ be the potential outcome with threshold $K$ and $Y(0)$ be the potential outcome without a threshold.
Assume that there are two types of customers: those who strategically exceed a specific threshold $K$~(\textit{i.e.,} bunching customers) and those who do not exceed the threshold~(\textit{i.e.,} non-bunching customers).
Bunching customers increase or adjust their spending to exceed the threshold and gain greater benefits than without it.
In contrast, non-bunching customers do not exceed the threshold or gain extra benefits.
Consequently, it is assumed that the bunching customers have potential outcomes $Y(t)$~(for $t \in \{0, 1\}$) which fall within the neighborhood of the threshold $K$, while non-bunching customers have the same potential outcomes, $Y(1)=Y(0)$ which lie outside the neighborhood.
Under the framework, the average treatment effect of the threshold on customers is defined as 
\begin{align}
\label{eq:causal-effect}
\Delta \equiv 
\mathbb{E} \left[Y(1) - Y(0) \mid  Y(t)\in N_K, \ t\in \{0,1\} \right],
\end{align}
where $N_K=[\underline{y}, \overline{y}]$ is the interval as the neighborhood of the threshold $K$.
This definition can be interpreted as the average treatment effect of the threshold on bunching customers, which is conceptually similar to the average treatment effect on the treated~(ATT) in the potential outcome framework.
In our application, we will deal with the situation with multiple thresholds, but we will describe our methodology under a single threshold for simplicity and discuss the extension in our application.

\subsection{Mixture Model for Bunching and Non-bunching Distributions}
\label{subsec:mixture-model}
Suppose we observe an independent and identically distributed observation, $Y_i$ for $i = 1, \ldots, n$, where $n$ is the number of customers. 
We assume that there are no bunching customers outside the neighborhood $N_K$ and they are not affected by the thresholds, namely $Y_i=Y_i(0)$ for $Y_i\in N_K^c$, where $N_K^c$ is the complement of $N_K$.
On the other hand, within the neighborhood $N_K$, bunching and non-bunching customers are mixed, that is, $Y_i=Z_iY_i(1)+(1-Z_i)Y_i(0)$, where $Z_i$ is an unobserved indicator of bunching~($Z_i=1$) and non-bunching~($Z_i=0$) customers. 
The key difference from typical causal inference frameworks is that the binary indicator $Z_i$ is unobserved. 
Although a precise estimation of the indicator $Z_i$ would be difficult, the ATT defined in (\ref{eq:causal-effect}) can be identified and estimated using a mixture model.

To perform a mixture model, we first assume the forms of distributions for $Y_i(t)$. 
Let $g(\cdot|\theta)$ and $f(\cdot|\gamma)$ be distributions for $Y_i(0)$ (non-bunching) and $Y_i(1)$ (bunching), respectively, where $\theta$ and $\gamma$ are unknown parameters. 
The bunching distribution $f(\cdot|\gamma)$ is expected to exhibit a high density within $N_K$, allowing us to assign a relatively simple model. 
In the numerical examples, we employ a skew-normal distribution~\citep{azzalini1985class}, as it offers the flexibility to model a density peak that may be slightly offset from the threshold $K$.
On the other hand, the support of the non-bunching distribution extends over the entire range of the observed data, so it is appropriate to adopt a flexible model. 
In our numerical illustration, we use the Singh-Maddala distribution~\citep{singh1976function}, which is widely recognized for its flexibility and excellent fit to income and expenditure data~(see Appendix A.3 for the density function and adjusted likelihood).

For the latent binary variable $Z_i$, we assume that $P(Z_i=1)=\pi=1-P(Z_i=0)$, where $\pi\in (0,1)$ is an unknown mixing parameter. 
This gives the generative distribution of $Y_i$ as 
\begin{equation}\label{eq:mixture}
p(Y_i|\Psi) = \pi f(Y_i|\gamma) + (1-\pi) g(Y_i|\theta).
\end{equation}
for $Y_i\in N_K$, where $\Psi=\{\pi, \theta,\gamma\}$ is a set of unknown parameters. 
Note that the distribution of $Y_i$ can be expressed as $p(Y_i| \Psi)=g(Y_i|\theta)$ for $Y_i\in N_K^c$.

Under the formulation~(\ref{eq:mixture}), the ATT can be expressed as 
\begin{equation}\label{eq:ATT}
\Delta(\Psi)\equiv 
\frac{\int_{y\in N_K} y f(y|\gamma)dy}{\int_{y\in N_K}f(y|\gamma)dy} - \frac{\int_{y\in N_K} y g(y|\theta)dy}{\int_{y\in N_K}g(y|\theta)dy}.
\end{equation}
The above quantity is a simple difference between the expectations of a bunching and a non-bunching distribution within $N_K$.  
Since we assume that most of the mass density of the bunching distribution $f(y|\gamma)$ is included in $N_K$, the first term of $\Delta(\Psi)$ will be approximately equal to the expectation of the density.

\subsection{Two-Step Bayesian Inference}
In our framework, the supports of bunching and non-bunching distributions are expected to differ, indicating that joint estimation of two distributions would be difficult. 
Therefore, we first estimate the non-bunching distribution $g(Y_i|\theta)$ from the set of observations $\mathcal{D}_{K^c}\equiv \{Y_i; Y_i\in N_K^c\}$, and fit the mixture model (\ref{eq:mixture}) with estimated $g(Y_i|\theta)$, using the set of observations $\mathcal{D}_{K}\equiv \{Y_i; Y_i\in N_K\}$. 
In what follows, we consider a Bayesian approach by assigning a prior distribution on unknown parameters. 
Such a Bayesian approach enables us to obtain point estimates and uncertainty measures, and is useful for handling heterogeneous treatment effects as considered in the subsequent section.

Regarding the first step, we obtain the posterior distribution of $\theta$ by assigning a prior distribution $\pi(\theta)$ for $\theta$. 
Since $g(Y_i|\theta)$ has a considerable mass outside $N_K^c$, the likelihood based on the sample $\mathcal{D}_K^c$ should be adjusted taking into account the probability falling into $N_K^c$.
We then consider the following posterior: 
\begin{equation}\label{eq:pos-theta}
p(\theta|\mathcal{D}_K^c)
\propto p(\theta)L(\theta; \mathcal{D}_K^c),
\end{equation}
where $L(\theta; \mathcal{D}_K^c)$ is the adjusted likelihood given by 
\begin{equation}\label{eq:ad-likelihood}
L(\theta; \mathcal{D}_K^c)
=\prod_{Y_i\in N_K^c}\frac{g(Y_i|\theta)}{\int_{y\in N_K^c} g(y|\theta)dy}.
\end{equation}
As the second step, we define the posterior of $(\pi,\gamma)$ as 
\begin{equation}\label{eq:pos-gamma}
\begin{split}
p(\pi,\gamma|\mathcal{D})
&=\int p(\pi, \gamma |\mathcal{D}_K, \theta) p^{\ast}(\theta|\mathcal{D}_K^c)d\theta\\
&=p(\pi, \gamma) \int \prod_{Y_i\in N_K} p(Y_i|\Psi)p^{\ast}(\theta|\mathcal{D}_K^c)d\theta,
\end{split}
\end{equation}
where $p(\pi, \gamma)$ is a joint prior for $(\pi, \gamma)$ and $p^{\ast}(\theta|\mathcal{D}_K^c)$ is a distribution to reflect the information from the posterior $p(\theta|\mathcal{D}_K^c)$. 
A possible example of $p^{\ast}$ is a Dirac measure on $\theta=\hat{\theta}$ with $\hat{\theta}$ being the posterior mean from (\ref{eq:pos-theta}), which will be used in our numerical examples.
The posterior computation of both (\ref{eq:pos-theta}) and (\ref{eq:pos-gamma}) can be performed by the Markov Chain Monte Carlo~(MCMC) algorithm using the probabilistic programming language \texttt{stan}~\citep{carpenter2017stan}, which takes advantage of the Hamiltonian Monte Carlo algorithm~\citep{duane1987hybrid} to sample complex probability distributions efficiently.
Given the posterior samples of $\Psi$, we can obtain posterior samples of $\Delta(\Psi)$ defined in (\ref{eq:ATT}), which give not only a point estimate but also a measure of uncertainty such as a $90\%$ credible interval.

\subsection{Theoretical Analysis}
Here, we discuss the theoretical justification of the proposed two-step Bayesian method in terms of posterior contraction. 
Let $\Delta_0$ be the true ATT.
We assume that the two models, $f(\cdot|\gamma)$ and $g(\cdot|\theta)$, are correctly specified, that is, there exists a true parameter $\Psi_0$ such that $\Delta_0=\Delta(\Psi_0)$. 
We also assume that $|\mathcal{D}_K|=\rho n$, namely $|\mathcal{D}_K^c|=(1-\rho) n$ for some $\rho\in (0,1)$, which means that the sample sizes in the neighborhood and the other region are not highly unbalanced. 

\medskip
\begin{theorem}
Assume regularity conditions in Appendix A.1.
Under $n \to \infty$, it holds that
$$
{\rm P}\bigl(|\Delta(\Psi)-\Delta_0|>Mn^{-1/2} \mid \mathcal{D}\bigr) \to 0,
$$
for a large constant $M$. 
\end{theorem}

\medskip
The proof is provided in Appendix A.2. 
Most of the regularity conditions are standard for the model (\ref{eq:mixture}) and adjusted likelihood (\ref{eq:ad-likelihood}), such as the identifiability of the two density functions, $f(\cdot|\gamma)$ and $g(\cdot|\theta)$, which is typically required to establish the posterior contraction rate \citep{ghosal2000convergence}.
The above theorem indicates that the proposed two-step Bayesian method holds posterior convergence to the true value at a rate $O(n^{-1/2})$. 
Hence, the information from the data can be successfully accumulated according to the sample size.
This property is essential to justify the hierarchical Bayesian formulation in Section~\ref{sec:HTE}.

\section{Heterogeneous Treatment Effect via Hierarchical Formulation}
\label{sec:HTE}

The advantage of the proposed Bayesian framework is its flexibility for model extensions, such as the estimation of heterogeneous treatment effects. 
Here, we discuss hierarchical Bayesian modeling to estimate heterogeneous treatment effects among customer subgroups.
Let customers be divided into $G$ subgroups according to their characteristics (e.g., age and sex), and let $Y_g(t), t\in \{0,1\}$ be the potential outcomes in the $g$th subgroup.
We assume that all groups share the same threshold $K$. 
Then, we can define ATT in each group, $\Delta_g$, in the same way as (\ref{eq:ATT}).
Let $\mathcal{D}^{(g)}=\{Y_{gi}, \ i=1,\ldots,n_g\}$ be a set of observations in the $g$th group.

We assume that the distributions of all the groups belong to the same family of distributions with different parameters, denoted by $f(\cdot|\gamma_g)$ and $g(\cdot|\theta_g)$.
Then, we consider the following generative distribution 
\begin{equation}
p_g(Y_{gi}|\Psi_g)= \pi_g f(Y_{gi}|\gamma_g) + (1-\pi_g)g(Y_{gi}|\theta_g),
\end{equation}
for $Y_{gi}\in N_K$, where $\pi_g$ is a heterogeneous mixing proportion. 
A direct application of the two-step approach given in the previous section to each group can suffer from instability when the group-level sample size, $n_g$, is small.  
Instead, we consider simultaneous estimation by introducing a random-effects structure $\theta_g\sim H_\theta(\alpha_{\theta})$ and $\gamma_g\sim H_\gamma(\alpha_{\gamma})$ with unknown parameters $\alpha_{\theta}$ and $\alpha_{\gamma}$. 
For $\pi_g$, we introduce a logit-normal distribution, namely $\log\{\pi_g/(1-\pi_g)\}\sim N(\mu_\pi, \sigma_{\pi}^2)$ with unknown $\mu_\pi$ and $\sigma_{\pi}^2$.
According to Theorem~1, the group-specific parameter $\Psi_g$ can be consistently estimated as $n_g$ increases, making more precise group-level information available. Consequently, the shrinkage effect induced by the random effect becomes weaker when $n_g$ is large, while it remains strong when $n_g$ is small. This behavior justifies using random effects, as it allows borrowing strength across groups adaptively depending on the sample size.

As the first step, the joint posterior of $\Theta=\{\theta_1,\ldots,\theta_G\}$ and $\alpha_\theta$ based on $\mathcal{D}_K^c=\{(\mathcal{D}_K^{(g)})^c, g=1,\ldots,G\}$ is given by 
\begin{equation}\label{eq:pos1-HB}
p(\Theta, \alpha_\theta|\mathcal{D}_K^c) \propto
p(\alpha_{\theta}) \prod_{g=1}^G h(\theta_g|\alpha_\theta) L(\theta_g; (\mathcal{D}_K^{(g)})^c),
\end{equation}
where $h(\theta_g|\alpha_\theta)$ is the density of $H_\theta(\alpha_{\theta})$. 
In the second step, the joint posterior of $\Psi=\{\Psi_1,\ldots,\Psi_G\}$ and $\alpha=\{\alpha_\gamma, \mu_\pi, \sigma_\pi^2\}$ is defined as 
\begin{equation}\label{eq:pos2-HB}
\begin{split}
p(\Psi, \alpha | \mathcal{D})
&\propto p(\alpha) \prod_{g=1}^G h_{\gamma}(\gamma_g|\alpha_\gamma)h_\pi(\pi_g|\mu_\pi, \sigma_\pi^2)\\
&\times \int \prod_{Y_{gi}\in N_K^{(g)}} p(Y_{gi}|\Psi_g)p^{\ast}(\Psi_g|D_K^c) d\theta_g,
\end{split}
\end{equation}
where $h_{\gamma}(\gamma_g|\alpha_\gamma)$ and $h_\pi(\pi_g|\mu_\pi, \sigma_\pi^2)$ are the density functions of $H_\gamma(\alpha_\gamma)$ and
${\rm logit}(\pi_g)\sim N(\mu_\pi, \sigma_\pi^2)$, respectively, and $p(\alpha)$ is a prior distribution.

In our simulation study and application, we use the following Singh-Maddala distribution~\citep{singh1976function} for the non-bunching distribution:
\begin{equation*}
g(Y_{gi}|\theta_g) = \frac{a_g q_g y^{a_g - 1}}{b_g^{a_g} (1 + (Y_{gi}/b_g)^{a_g})^{q_g+1}}, 
\end{equation*}
where $\theta_g=(a_g, b_g, q_g)$, $a_g$ and $q_g$ are shape parameters and $b_g$ is a scale parameter.
For these parameters, we assume log-normal random effects, $\log (a_g)\sim N(\mu_a, \sigma_a^2)$, $\log (b_g)\sim N(\mu_b, \sigma_b^2)$ and $\log (q_g)\sim N(\mu_q, \sigma_q^2)$, with unknown parameters, $\alpha_\theta=(\mu_a, \sigma_a^2, \mu_b, \sigma_b^2, \mu_q, \sigma_q^2)$. 
For the bunching distribution, we employ the skew-normal distribution 
\begin{equation*}
f(Y_{gi}|\gamma_g) = \frac{2}{\omega_g}\phi\left(\frac{Y_{gi}-\beta_g}{\omega_g}\right)\Phi\left(\delta_g \frac{Y_{gi}-\beta_g}{\omega_g}\right), 
\end{equation*}
where $\gamma_g=(\beta_g, \omega_g, \delta_g)$, which are location, scale, and skewness parameters, respectively, $\phi(\cdot)$ and $\Phi(\cdot)$ are the density and cumulative distribution functions of $N(0,1)$.
For the group-level parameters, we employ random effects models, $\beta_g\sim N(\mu_\beta, \sigma_\beta^2)$, $\log (\omega_g) \sim N(\mu_\omega, \sigma_\omega^2)$ and $\delta_g\sim N(\mu_\delta, \sigma_\delta^2)$ with unknown parameters $\alpha_\gamma=(\mu_\beta, \sigma_\beta^2, \mu_\omega, \sigma_\omega^2, \mu_\delta, \sigma_\delta^2)$.
The detailed settings of the prior distributions are provided in Appendix B. 
We can again use \texttt{stan} to generate random samples from the posterior distributions (\ref{eq:pos1-HB}) and (\ref{eq:pos2-HB}).
Based on the group-level parameters, $\Psi_g$, we can define the group-wise heterogeneous treatment effect $\Delta(\Psi_g)$ in the same way as (\ref{eq:ATT}), and the posterior distributions of $\Delta(\Psi_g)$ can be approximated by the posterior samples of $\Psi_g$. 
This gives point estimates and measures of uncertainty quantification of $\Delta(\Psi_g)$.

\section{Simulation Studies}
We demonstrate the effectiveness of our proposed methods by comparing their performance with the standard RDD method.
We assume that the customer threshold $K$ is known and set to 50, and the observed data originate from a population structured into $G=100$ distinct subgroups.
The observed data for each subgroup $g~(g=1, \ldots, 100)$ are generated by a mixture model consisting of a bunching distribution and a non-bunching distribution, weighted by subgroup-specific proportions $\pi_g$ and $1-\pi_g$, respectively.
The parameters defining the distributions of components, including the proportions, can vary from subgroup to subgroup.
We equally divided $G$ groups into four clusters and set the same group-specific sample sizes $n_g$~($g=1, \ldots, G$) to the same values within the same clusters.
The sample sizes of the clusters are set to (50, 100, 200, 300).
We consider two scenarios to evaluate the model performance under different conditions.
Scenario~(A) assumes moderate bunching with low heterogeneity across subgroups.
In contrast, Scenario~(B) presents a more sparse bunching and higher heterogeneity than Scenario~(A).
The detailed simulation settings are deferred to Appendix C.

For the simulated dataset, we fitted the proposed BMTM and HBMTM models based on 12,000 posterior samples consisting of four chains each with 3,000 samples after discarding the first 3,000 samples as burn-in. 
We use \texttt{CmdStanPy}~(version 1.2.5) for model fitting.
The settings for the prior distributions of BMTM and HBMTM are provided in Appendix B.
Based on the posterior samples, we computed the posterior means and the $90\%$ highest density interval~(HDI) of the group-wise ATT. 
For comparison, we also applied an RDD method, which estimates the effect of treatment by measuring the discontinuity in the probability density of $Y$ at the threshold $K$.
This can be achieved by applying the kernel density estimation separately on either side of $K$ and computing the difference between the density values. 
In our implementation, a boundary correction technique for density estimation~\citep{jones1993simple} was applied, and a bandwidth of 10, centered at the threshold $K=50$ (covering the range $[40, 60]$), was used.
Note that this RDD does not provide interval estimates.

The performance of the point estimation is evaluated using Mean Absolute Error~(MAE), defined as
\begin{align*}
  \text{MAE} &= \frac{1}{G} \sum_{g=1}^{G} \left |\Delta_g - \widehat{\Delta}_g\right|,
\end{align*}
Regarding the interval estimation, we use the Coverage Probability~(CP), Average Length~(AL), and Interval Score~(IS)~\citep{gneiting2007strictly}, defined as
\begin{align*}
    \text{CP} &= \frac{1}{G} \sum_{g=1}^{G} \mathbf{1}(l_g \leq \Delta_g \leq u_g),\ \  \text{AL} = \frac{1}{G} \sum_{g=1}^{G} (u_g - l_g), \\
    \text{IS} &= \frac{1}{G} \sum_{g=1}^{G} \biggl\{ (u_g -l_g) + \frac{2}{\alpha} (l_g - \Delta_g) \mathbf{1}(\Delta_g < l_g) \\
              & \quad + \frac{2}{\alpha} (\Delta_g - u_g) \mathbf{1}(\Delta_g > u_g) \biggr\},
\end{align*}
where $l_g$ and $u_g$ are the lower and upper bounds of $100(1 - \alpha)$\% HDI for each subgroup $g$, respectively, and $\mathbf{1}(\cdot)$ is the indicator function.
Note that $\alpha=0.1$ in our settings. 
The CP measures the proportion of true values covered by the interval, with an ideal value of $1 - \alpha$.
The IS is a composite metric that jointly evaluates the AL and CP of the interval.
The IS increases when the prediction interval generated by the model is too broad and increases when the observed value falls outside of this interval.
Table~\ref{tab:simulation-hierarchical} summarizes the above performance measures averaged over $100$ Monte Carlo replications. 

\begin{table}[tb]
\centering
\begin{tabular}{cccccccc}
\hline
Scenario  & Method & MAE & CP & AL & IS \\ \hline
    & RDD   & 3.31 & -- & -- & --  \\
(A) & BMTM  & 0.78 & 0.91 & 4.03 & 4.65  \\
    & HBMTM & \textbf{0.33} & 0.84 & 1.20  & \textbf{1.85} \\ 
\hline
    & RDD   & 3.50 & -- & -- & --  \\
(B) & BMTM  & 1.79 & 0.94 & 9.05 & 9.88  \\
    & HBMTM & \textbf{0.37} & 0.88 & 1.62 & \textbf{2.15}\\ 
\hline
\end{tabular}
\caption{Simulation results of the proposed and baseline methods in the subgroup setting. The minimum values of MAE and IS are shown in bold.}
\label{tab:simulation-hierarchical}
\end{table}
In both scenarios, our proposed methods, BMTM and HBMTM, substantially outperform the baseline RDD.
In particular, HBMTM consistently demonstrates superior performance with the lowest MAE and IS, and the superiority is evident in Scenario~(B). 
In Scenario~(B) with a weaker bunching signal, the accuracy of BMTM degrades~(MAE increases from 0.78 to 1.79), while that of HBMTM remains remarkably stable and accurate. 
In Scenario~(A), although CP of BMTM is slightly closer to the nominal 90\% than that of HBMTM, HBMTM maintains reliable CP across both scenarios.
These findings demonstrate that the hierarchical structure of HBMTM successfully borrows strength across subgroups to deliver highly accurate and precise causal effect estimates, even when the identifying bunching signal is weak.

\begin{figure}
  \centering
  \includegraphics[width=\columnwidth]{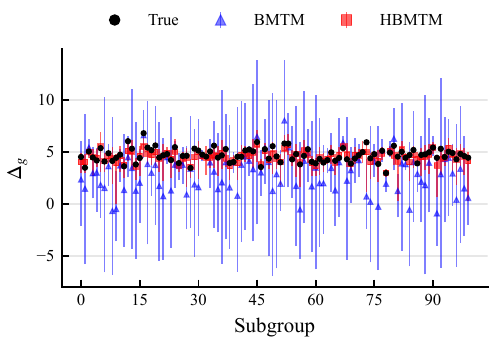}
  \caption{Point estimates and 90\% credible intervals of the subgroup-specific causal effects $\Delta_g$ by BMTM and HBMTM.}
  \label{fig:comparison-subgroups-hierarchical}
\end{figure}

Figure~\ref{fig:comparison-subgroups-hierarchical} visually contrasts the subgroup-specific causal effects estimates $\Delta_g$ from BMTM and HBMTM in Scenario~(B).
It confirms that the hierarchical model, HBMTM, consistently provides more accurate point estimates and narrower 90\% credible intervals.
These visual comparisons further emphasize the effectiveness of our proposed HBMTM method in estimating the threshold's causal effect on customers, even in the presence of small sample sizes within subgroups.

\section{Application to Promotion with Thresholds in Marketing}
In this section, we apply the HBMTM to a real-world dataset from a marketing promotion in which consumers could earn incentives by exceeding spending thresholds.
A month-long promotion offered consumers entry into a prize draw for spending more than 30,000 yen.
The winning odds were tripled for spending more than 50,000 yen, multiplied by five for spending more than 70,000 yen, and the winners were announced after the promotion.

\begin{figure}
  \centering
  \includegraphics[width=\columnwidth]{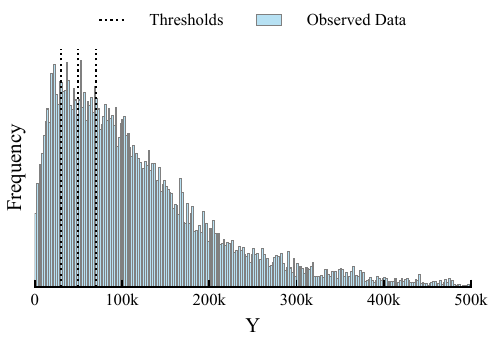}
  \caption{Observed distribution of the total payment amount during the promotion~(skyblue) and the three thresholds~(red).}
  \label{fig:actual-histogram}
\end{figure}

Figure~\ref{fig:actual-histogram} shows the histogram of the total payment amount, $Y$, per participant during the promotion, with the thresholds indicated by red lines.
As shown in Figure~\ref{fig:actual-histogram}, the distribution of $Y$ is asymmetric, with its peak skewed to the left and a long and heavy tail extending to the right.
Such a right-skewed, heavy-tailed distribution is a typical characteristic of expenditure data.
We then adopt the Singh-Maddala distribution for non-bunching distributions, as explained in Section~\ref{sec:HTE}, a flexible three-parameter distribution for positive-valued data.

In this study, we defined subgroups based on total spending in the previous month to investigate how the impact of promotional thresholds varies across customers with different baseline spending habits. 
Defining subgroups based on prior spending is motivated by the premise that a customer's proximity to the promotional thresholds is a key determinant of their behavioral response.
Specifically, we constructed $G=21$ subgroups by grouping participants into 10,000 yen intervals based on their previous month's spending, with the tier exceeding 200,000 yen treated as a single, top-coded group.

We applied our proposed method, HBMTM, to this promotion dataset.
The detailed settings for the prior distributions of HBMTM are provided in Appendix B.2.
Since the dataset has three threshold values, denoted by $K_1=30,000, K_2=50,000$ and $K_3=70,000$, we first define $\mathcal{D}_{K^c}$ by excluding observations that fall within the interval $[20,000, 80,000]$, which represents the combined range of $\pm$ 10,000 yen around each of the three thresholds.
Then we separately apply the mixture model (\ref{eq:mixture}) to each neighborhood, $N_{K_m}=[K_m-10,000, K_m+10,000]$ for $m=1,2,3$. 
Here, we assume that the three thresholds are mutually independent, meaning that the presence of the others does not influence customers around one threshold.

\begin{figure}
  \centering
  \includegraphics[width=\columnwidth]{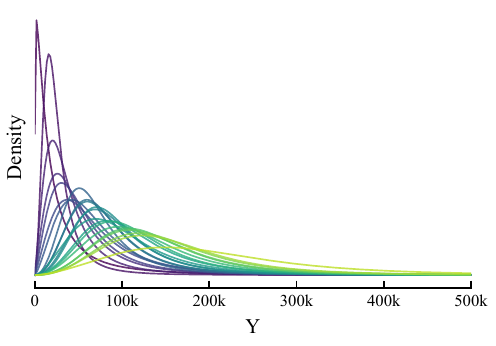}
  \caption{The non-bunching distribution for each subgroup, estimated using the Singh-Maddala distribution.}
  \label{fig:sm-non-bunching}
\end{figure}

Figure~\ref{fig:sm-non-bunching} shows the Singh-Maddala distribution for each subgroup estimated in the first step of HBMTM.
As shown in Figure~\ref{fig:sm-non-bunching}, the distributional shapes differ significantly among subgroups. 
This difference in shape suggests the existence of significant spending heterogeneity, which in turn validates our adoption of the highly flexible Singh-Maddala distribution. 

\begin{figure*}[!t]
\centering
\subfloat[Threshold at 30,000]{\includegraphics[width=0.48\textwidth]{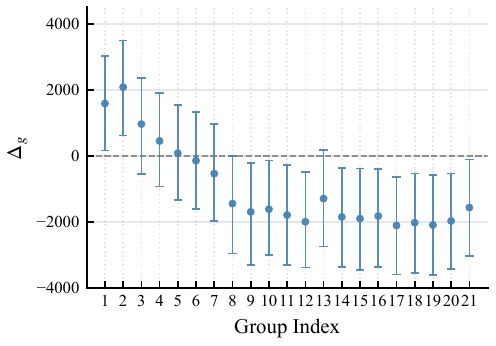} \label{fig:th_30000}}
\hfill
\subfloat[Threshold at 50,000]{\includegraphics[width=0.48\textwidth]{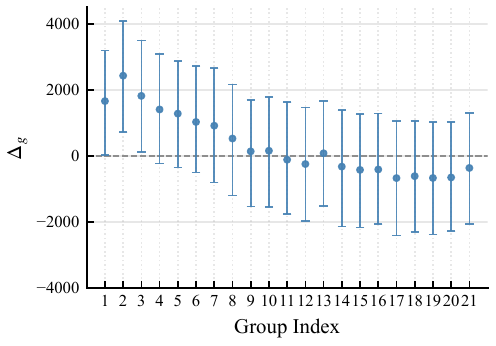} \label{fig:th_50000}}
\centering
\subfloat[Threshold at 70,000]{\includegraphics[width=0.48\textwidth]{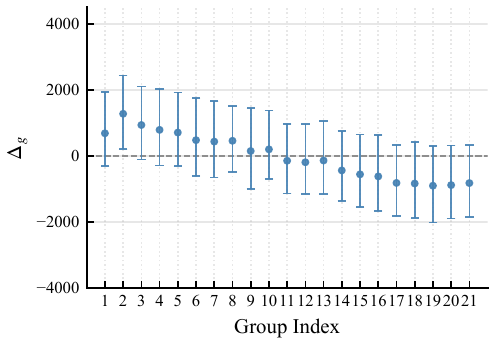} \label{fig:th_70000}}
\caption{Point estimates and 90\% credible intervals of each subgroup's causal effect $\Delta_g$ obtained by the proposed method HBMTM at each threshold.}
\label{fig:application-results}
\end{figure*}

Figure~\ref{fig:application-results} provides the posterior median and 90\% HDI for each subgroup's causal effect $\Delta_g$ across the three thresholds. 
The results presented in Figure~\ref{fig:application-results} demonstrate a significant heterogeneity in causal effects across subgroups, with a consistent trend observed across all thresholds.
Specifically, a clear positive causal effect was observed at every threshold in subgroups where the previous month's spending was below or slightly above the threshold~(e.g., Group index 1-4 for K1).
For some groups, 90\% HDI was entirely within the positive range, suggesting statistical confidence in the effect.
This positive effect aligns with the intuitive hypothesis that consumers increase spending to achieve the incentive, a finding consistent with the hypotheses suggested in existing marketing research~\citep[e.g.][]{kivetz2006goal}.

In contrast, a consistent trend of diminishing and eventually negative causal effects was observed as spending in the previous month significantly exceeded each threshold~(e.g., Group index $\geq$ 6 for K1).
This spending-dampening effect on high-spending subgroups may be explained by the anchoring effect~\citep[e.g.][]{holst2015anchoring, zong2022experimental}.
It is plausible that the relatively low promotional thresholds acted as a new psychological anchor for consumers, leading those with higher baseline spending levels to adjust their expenditures downward, influenced by this reference point.
Although more research is needed to elucidate the detailed mechanisms underlying these effects, these results strongly suggest that our proposed method effectively captures heterogeneous behavioral changes under threshold manipulation.

\section{Concluding Remarks}
This study addresses the challenge of estimating the causal effects of marketing thresholds under strategic customer manipulation, a scenario where standard RDD and bunching estimation methods face significant limitations.
To overcome this issue, we propose BMTM, a novel Bayesian mixture model framework. 
This framework robustly estimates causal effects by modeling the observed spending distribution as a mixture of two latent distributions: one representing customers unaffected by the threshold and another representing those strategically affected.
Furthermore, we introduce its hierarchical extension, HBMTM, which, to our knowledge, is the first method to provide stable and accurate estimates of heterogeneous causal effects across customer subgroups, even with small sample sizes.
Our empirical results from both simulations and a real-world case study demonstrate the effectiveness of our methods, yielding more accurate and reliable estimates than conventional approaches.
This study provides a practical tool for data-driven marketing and contributes a principled approach for causal inference under strategic manipulation. 
Future research could explore more flexible non-parametric models within this framework and extend its application to other domains where agents respond to algorithmic or policy-based thresholds.

\bibliography{aaai2026}

\appendix
\clearpage

\renewcommand{\thefigure}{A\arabic{figure}}
\renewcommand{\thetable}{A\arabic{table}}

\section{Technical Details}

\subsection{Regularity conditions}
We introduce the regularity conditions required for the proof of Theorem 1.
In what follows, we define ${\rm KL}(p(x)\|q(x))$ as the Kullbuck-Leibler divergence between two density functions, $p(x)$ and $q(x)$, defined as 
$$
{\rm KL}(p(x)\|q(x)) = \int p(x)\log\left\{\frac{p(x)}{q(x)}\right\}dx
$$
We first define the adjusted likelihood for a single observation as 
$$
L(x|\theta)= \frac{g(x|\theta)}{\int_{y\in N_K^c} g(y|\theta)dy}.
$$
To establish the posterior contraction rate given in Theorem~1, we assume the following regularity conditions: 
\begin{itemize}
\item[(C1)] 
The parameter space of $\Phi$ is compact and the true $\Phi_0=\{\pi_0, \theta_0, \gamma_0\}$ is an interior point of the space. 

\item[(C2)] 
The adjusted likelihood function $L(x|\theta)$, non-bunching distribution $g(x|\theta)$, and the bunching distribution $f(x|\gamma)$ are second-order continuously differentiable. 

\item[(C3)]
For all $\gamma\neq \gamma_0$ and $\theta\neq \theta_0$, it holds that ${\rm KL}(g(x|\theta_0),g(x|\theta))>0$ and ${\rm KL}(f(x|\gamma_0),f(x|\gamma))>0$.

\item[(C4)]
The Fisher information matrices 
$$
E\left[\frac{\partial \log g(Y|\theta)}{\partial \theta}\frac{\partial \log g(Y|\theta)}{\partial \theta^\top }\right]
$$
$$
E\left[\frac{\partial \log f(Y|\gamma)}{\partial \gamma}\frac{\partial \log f(Y|\gamma)}{\partial \gamma^\top }\right]
$$
are both positive definite. 

\item[(C5)]  
There exist $\delta>0$ such that a prior distribution for $\Psi$ satisfies $\inf_{\|\Psi-\Psi_0\|<\delta}\pi(\Psi)>0$.

\item[(C6)]  
For any $\gamma$ and $\theta$, ${\rm KL}(f(x|\gamma)\|g(x|\theta))>0$.

\item[(C7)]
The surrogate posterior $p^{\ast}(\theta \mid \mathcal{D}_K^c)$ is a continuous mapping of the original posterior $p(\theta \mid \mathcal{D}_K^c)$ with respect to the topology of weak convergence.

\item[(C8)]
$\tau(\Psi)$ is a continuously differentiable function of $\Psi$.
\end{itemize}

Conditions (C1)-(C5) are the standard conditions for establishing the contraction rate of posterior distributions \citep{van2000asymptotic,ghosal2000convergence}.
Condition (C6) ensures that the mixture model is identifiable.
Condition (C7) is satisfied by, for example, setting $p^{\ast}$ to a Dirac measure on the posterior mean, as used in our simulation and application. 
Condition (C8) is needed to establish posterior convergence given the convergence of the posterior of $\Psi$.

In our simulation and application, the Singh-Madala and skew-normal distributions for $g(y|\theta)$ and $f(y|\gamma)$, respectively. 
The properties of the Fisher information matrices of these distributions have been investigated in \citet{mahmoud2015fisher} for the Singh-Madala distribution and in \citet{franco2024information} for the skew-normal distribution.

\subsection{Proof of Theorem 1}
Under conditions (C1)-(C5), the posterior distribution $p(\theta|\mathcal{D}_K^c)$ holds posterior contraction
$$
P(\|\theta-\theta_0\|\geq M_1n^{-1/2} \mid \mathcal{D}_K^c)\to 0
$$
as $n\to\infty$ for a constant $M_1>0$ \citep[e.g.][]{ghosal2000convergence}.
Then, we also obtain the same posterior contraction result for the surrogate posterior $p^{\ast}(\theta|\mathcal{D}_K^c)$ from (C7).

Given the true value of $\theta_0$, the mixture model is expressed as 
$$
p(y|\pi,\gamma)=\pi f(y|\gamma) + (1-\pi) g(y|\theta_0).
$$
It holds that 
$$
a(y|\pi,\gamma)\equiv 
\frac{\partial \log p(y|\pi,\gamma)}{\partial \pi}
=\frac{f(y|\gamma)-g(y|\theta_0) }{p(y|\pi,\gamma)}
$$
$$
b(y|\pi,\gamma)\equiv 
\frac{\partial \log p(y|\pi,\gamma)}{\partial \gamma}
=\frac{ \pi \cdot \partial f(y|\gamma)/\partial\gamma  }{p(y|\pi,\gamma)}
$$
Then, the Fisher information matrix $I(\pi, \gamma)$ is given by
$$
I(\pi, \gamma) = \left[
\begin{array}{cc}
I_{\pi\pi} & I_{\pi\gamma}\\
I_{\pi\gamma}^\top  & I_{\gamma\gamma}
\end{array}
\right],
$$
where 
\begin{align*}
&I_{\pi\pi}= E[ a(Y|\pi,\gamma)^2], \ \ I_{\pi\gamma}=E[a(Y|\pi,\gamma)b(Y|\pi,\gamma)],\\ 
&I_{\gamma\gamma}=E[b(Y|\pi,\gamma)b(Y|\pi,\gamma)^\top.
\end{align*}
To state that $I(\pi, \gamma)$ is positive definite, it is sufficient to show that (i) $I_{\gamma\gamma}$ is positive definite and (ii) $ I_{\pi\pi} - I_{\pi\gamma} I_{\gamma\gamma}^{-1} I_{\pi\gamma}^\top>0$.   
For (i), $I_{\gamma\gamma}$ can be expressed as 
$$
I_{\gamma\gamma}=E\left[\left(\frac{\pi f(Y|\gamma)}{p(Y|\pi,\gamma)}\right)^2\cdot \frac{\partial \log f(Y|\gamma)}{\partial \gamma} \cdot \frac{\partial \log f(Y|\gamma)}{\partial \gamma^\top}\right].
$$
From (C4), it holds that $E[(c^\top \partial \log f(Y|\gamma)/\partial \gamma)^2]>0$ for an arbitrary non-zero vector $c$. 
This indicates that the function $c^\top \partial \log f(y|\gamma)/\partial \gamma$ is not identical to $0$.
Hence, we have 
$$
c^\top I_{\gamma\gamma}c
=E\left[\left(\frac{\pi f(Y|\gamma)}{p(Y|\pi,\gamma)}\right)^2 \cdot \left(c^\top \frac{\partial \log f(Y|\gamma)}{\partial \gamma} \right)^2\right]>0.
$$
Moreover, the condition (ii) follows from the Cauchy-Schwarz inequality and the fact that $a(y|\pi,\gamma)$ and $b(y|\pi,\gamma)$ are not linearly dependent under the identifiability condition (C6). 
Hence, the Fisher information matrix $I(\pi, \gamma)$ is positive definite, which leads to the posterior contraction of the conditional posterior distribution $p(\pi,\gamma|\mathcal{D}_K, \theta_0)$
$$
P(\|(\pi,\gamma)-(\pi_0,\gamma_0)\|\geq M_2n^{-1/2})\to 0
$$
as $n\to\infty$ for a constant $M_2>0$.

Let $A_n=\{(\pi, \gamma) \mid \|(\pi, \gamma) - (\pi_0, \gamma_0)\| > M_2 n^{-1/2} \}$. 
Let $p(A_n | \mathcal{D})$ be the marginal posterior probability of $A_n$, and let $p(A_n | \mathcal{D}_K, \theta)$ be its conditional posterior probability.
It holds that 
\begin{align*}
p(A_n | \mathcal{D})
&= \int p(A_n | \mathcal{D}_K, \theta) p^{\ast}(\theta |\mathcal{D}_K^c) d\theta \\
&= \int_{\|\theta - \theta_0\| \leq \epsilon_n}p(A_n | \mathcal{D}_K, \theta) p^{\ast}(\theta |\mathcal{D}_K^c)  d\theta\\
&+ \int_{\|\theta - \theta_0\| >\epsilon_n}p(A_n | \mathcal{D}_K, \theta) p^{\ast}(\theta |\mathcal{D}_K^c) d\theta,
\end{align*}
where $\epsilon_n=M_1n^{-1/2}$.
The second term is bounded by the posterior probability of the event $\|\theta - \theta_0\| > \epsilon_n$, which goes to $0$ as $n\to\infty$ due to the posterior contraction of $p^{\ast}(\theta |\mathcal{D}_K^c)$.
For the first term, we first note that 
$$
\sup_{\|\theta - \theta_0\| \leq \epsilon_n} |p(y|\pi, \gamma, \theta) - p(y|\pi, \gamma, \theta_0)|\leq C_1 \epsilon_n,
$$
for a constant $C_1>0$ and arbitrary $y$, $\pi$ and $\theta$, from (C1) and (C2).
This leads to 
$$
\sup_{\|\theta - \theta_0\| \leq \epsilon_n} |p(\pi,\gamma | \mathcal{D}_K, \theta) - p(\pi,\gamma | \mathcal{D}_K, \theta_0)|\leq C_2 \epsilon_n
$$
for a constant $C_2>0$ and arbitrary $\pi$ and $\theta$.
Then, it follows that 
\begin{align*}
&\int_{\|\theta - \theta_0\| \leq \epsilon_n} p(A_n | \mathcal{D}_K, \theta) \pi^{\ast}(\theta | \mathcal{D}_K^c) d\theta\\
&\leq \int_{\|\theta - \theta_0\| \leq \epsilon_n} |p(A_n | \mathcal{D}_K, \theta) - p(A_n | \mathcal{D}_K, \theta_0)|\pi^{\ast}(\theta | \mathcal{D}_K^c) d\theta\\
&\ \ \ \  + \int_{\|\theta - \theta_0\| \leq \epsilon_n} p(A_n | \mathcal{D}_K, \theta_0)|\pi^{\ast}(\theta | \mathcal{D}_K^c) d\theta \\
& \leq C_2 \varepsilon_n + p(A_n | \mathcal{D}_K, \theta_0), 
\end{align*}
which goes to $0$ as $n\to\infty$, from the posterior contraction of the conditional posterior. 
Therefore, it holds that $\pi(A_n | \mathcal{D})\to 0$ as $n\to\infty$, which completes the proof from (C8).

\subsection{Singh-Madala distribution and its adjusted likelihood}

The Singh-Madala distribution is a flexible three-parameter distribution on the positive real line, often used to model economic data such as income and expenditure.
The parameter vector is denoted as $\theta = (a, b, q)$, where $a$ and $q$ are shape parameters, and $b$ is a scale parameter.
The probability density function is given by
\begin{equation*}
g(y|\theta) = \frac{a q y^{a - 1}}{b^a (1 + (y/b)^a)^{q+1}}, \ \ y > 0.
\end{equation*}
In addition, its distribution function is given by 
$$
G(y|\theta) = 1 - \left(1 + \left(\frac{y}{b}\right)^a\right)^{-q}, \quad y > 0.
$$
Hence, the adjusted likelihood function under $N_K=[K-a, K+a]$ for some $a\in (0, K)$ is obtained as 
\begin{align*}
L(\theta; \mathcal{D}_K^c)=\prod_{Y_i\in \mathcal{D}_K^c} \frac{g(Y_i|\theta)}{1-G(K+a|\theta)+G(K-a|\theta)}.
\end{align*}

\section{Prior Specifications}

\subsection{Prior distributions for BMTM and HBMTM in simulation studies} 
For the first step model of BMTM, the prior for each element of the unknown parameter vector $\theta = (a, b, q)$ is specified as follows:
\begin{gather*}
\log(a), \log(b) \sim N(0, 1.5^2), \ \ \ q \sim N(\log(40), 1).
\end{gather*}
For the second step model of BMTM, the priors for the unknown parameters $\gamma = (\beta, \omega, \delta)$ and $\pi$ are specified as follows:
\begin{gather*}
\omega \sim N(0, 10^2), \ \ \delta \sim N(0, 2^2), \ \ 
\log \left\{ \frac{\pi}{1- \pi} \right\} \sim N(0, 1.5^2).
\end{gather*}
For the location parameter $\beta$ of the skew-normal distribution, we set $\beta = K$.
This specification reflects the prior belief that the location of the bunching distribution is at the threshold $K$.

For HBMTM, the priors for the unknown parameter $\alpha_{\theta} = (\mu_{a}, \sigma_a^2, \mu_b, \sigma_b^2, \mu_q, \sigma_q^2)$ in the first step model are specified as follows:
\begin{gather*}
\mu_a \sim N(0, 2.5^2) , \ \ 
\mu_b \sim N(3, 2^2), \ \
\mu_q \sim N(0, 2.5^2), \\
\sigma_a, \sigma_b, \sigma_q \sim N^{+}(0, 1),
\end{gather*}
where $N^{+}(\cdot)$ denotes a half-normal distribution, a truncated normal distribution with a lower bound of 0.
The priors for the unknown parameter in the random effects distributions for the heterogeneous mixing proportion $\pi_g$ and the skew-normal parameters are specified as follows:
\begin{gather*}
\mu_{\omega} \sim N(2, 1), \ \
\mu_{\delta} \sim N(0, 1), \ \
\mu_{\pi} \sim N(0, 1.5^2), \\
\sigma_{\omega}, \sigma_{\delta}, \sigma_{\pi} \sim N^{+}(0, 1).
\end{gather*}
For the heterogeneous location parameter $\beta_g$, we set its hyperparameters to $\mu_\beta = K$ and $\sigma_\beta = 0$, which implies that $\beta_g=K$ for all $g=1,\ldots, G$.
This choice reflects the prior belief that the location of the bunching distribution for every group is at the threshold $K$.

\subsection{Prior distributions for HBMTM in real-world application}
\label{subsec:prior-application}
For the first step model, the priors for the unknown parameters $\alpha_\theta=(\mu_a, \sigma_a^2, \mu_b, \sigma_b^2, \mu_q, \sigma_q^2)$ are specified as follows:
\begin{gather*}
\mu_a \sim N(1, 1), \ \ 
\mu_b \sim N(10, 2^2), \ \ 
\mu_q \sim N(0, 1), \\ 
\sigma_a \sim N^{+}(0.5, 1), \ \ 
\sigma_b \sim N^{+}(10, 2^2), \ \ 
\sigma_q \sim N^{+}(0, 1).
\end{gather*}
For the second step model, the priors for the unknown parameters of the random effects distributions for the heterogeneous mixing proportion $\pi_g$ and the skew-normal parameters are specified as follows: 
\begin{gather*}
\mu_{\pi} \sim N(2, 1), \ \ 
\mu_{\omega} \sim N(7, 0.5^2), \ \ 
\mu_{\delta} \sim N(0, 1), \\
\sigma_{\pi} \sim N^{+}(0, 0.5^2), \ \ 
\sigma_{\omega} \sim N^{+}(0, 1), \ \ 
\sigma_{\delta} \sim N^{+}(0, 0.5^2).
\end{gather*}
As in the simulation, we set the parameters for the heterogeneous location parameter $\beta_g$ to $\mu_\beta = K$ and $\sigma_\beta = 0$, which implies that $\beta_g = K$ for all $g = 1, \ldots, G$.
This specification reflects the prior belief that the location of the bunching distribution is at the threshold $K$.

\section{Supplementary Details of Numerical Studies }

\subsection{Details of data generating process in simulation studies} 
The data for our simulation studies are generated from the hierarchical model described in Section~4.
The parameter settings for the bunching and non-bunching distributions, which are common to both scenarios (A) and (B), are as follows:
\begin{gather*}
\omega_g \sim N(\mu_{\omega}, \sigma_{\omega}^2), \ \ 
\delta_g \sim N(\mu_\delta, \sigma_{\delta}^2), \ \ 
a_g \sim N^{+}(\mu_a, \sigma_a^2), \\ 
b_g \sim N^{+}(\mu_b, \sigma_b^2), \ \ 
q_g \sim N^{+}(\mu_q, \sigma_q^2), \ \ 
\end{gather*}
where the hyperparameters are drawn from:
\begin{gather*}
\mu_{\omega} \sim N(3, 0.1^2), \ \ 
\mu_{\delta} \sim N(4, 0.1^2), \ \ 
\mu_a \sim N(3.5, 0.1^2), \\ 
\mu_b \sim N(39, 1), \ \ 
\mu_q \sim N(1.5, 0.1^2), \ \ 
\sigma_b \sim N^{+}(2, 1), \\
\sigma_{\omega}, \sigma_{\delta} \sim N^{+}(0.5, 0.1^2), \ \
\sigma_a, \sigma_q \sim N^{+}(0.2, 0.1^2),
\end{gather*}
The location parameter for the bunching, $\beta_g$, is fixed in the threshold $K$ for all groups.

The two scenarios differ only in the heterogeneous mixing proportion $\pi_g$.
Assuming that $\log \{ \pi_g / (1 - \pi_g)\} \sim N(\mu_{\pi}, \sigma_{\pi}^2)$, the settings for each scenario are as follows: 
\begin{gather*}
\text{Scenario~(A):} \ \ \mu_{\pi} \sim N(-2, 0.1^2), \ \ \sigma_{\pi} \sim N^{+}(0.5, 0.1^2), \\
\text{Scenario~(B):} \ \ \mu_{\pi} \sim N(-4, 0.1^2), \ \ \sigma_{\pi} \sim N^{+}(1.5, 0.1^2).
\end{gather*}
Scenario~(A) represents a case where the mixing proportion of the bunching component is moderate and exhibits low heterogeneity across groups.
In contrast, Scenario~(B) features a more sparse bunching component~(\textit{i.e.,} low mixing proportion) but with higher heterogeneity than Scenario~(A).
By comparing these two scenarios, we examine how the model's performance is affected by the overall share of the bunching component and the degree of its cross-group heterogeneity.

\subsection{Validation of the bunching region}
To validate this model specification, we verify that the estimated bunching distribution has a negligible density outside this region.
Our proposed method assumes that the bunching is fully contained within the bunching region $N_{K_m} = [K_m - 10,000, K_m + 10,000]$ for $m = 1, 2, 3$.
Specifically, we confirm that the density is zero at the endpoints of the interval, $K_m \pm 10,000$.

\begin{figure}[!t]
    \centering
    \subfloat[Lower bound]{
        \includegraphics[width=0.8\columnwidth]{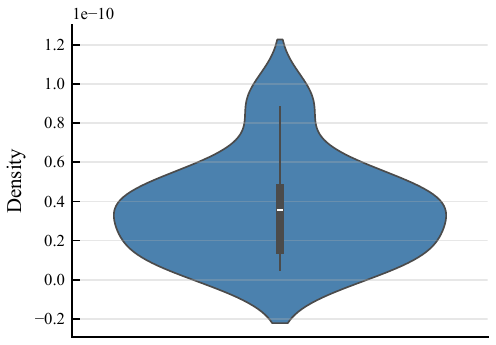}
        \label{fig:density-lower}
    }
    
    \subfloat[Upper bound]{
        \includegraphics[width=0.8\columnwidth]{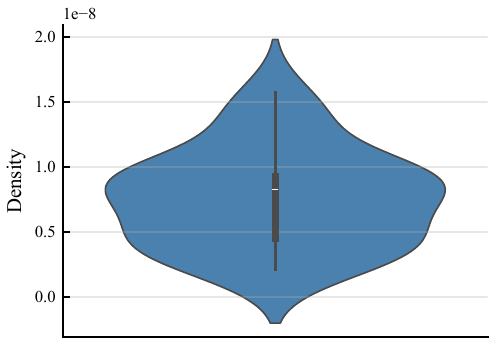}
        \label{fig:density-upper}
    }
    \caption{Distribution of bunching region $N_{K_m}$ endpoints across subgroups (Threshold: 50,000). 
    Violin plots show the estimated (a) lower and (b) upper bounds of the region.}
    \label{fig:density-zero}
\end{figure}

We present the case for $K_2 = 50,000$ as a representative example.
Specifically, for the bunching distribution estimated for each subgroup, we calculated its probability density at the two endpoints of $N_{K_2}$: 40,000 and 60,000.
Figure~\ref{fig:density-zero} shows violin plots of these density values at two endpoints across all subgroups.
As shown in Figure~\ref{fig:density-zero}, the distributions of the density values in both (a) the lower and (b)the upper bounds are concentrated at nearly zero.
These results indicate that the estimated bunching distribution has negligible leakage outside the boundaries of the bunching region $N_{K_2}$. 
Therefore, these results support the model's assumption that the bunching effect is limited to the interval $[K_m - 10000, K_m + 10,000]$, thus justifying the model specification.

\subsection{Sensitivity analysis on the prior in real-world application}
In this section, we present a sensitivity analysis to assess the robustness of our findings to the prior specification for the bunching location parameter, $\beta_g$.
Our primary analysis~(See Section~6) fixed this parameter at the thresold for all groups~(\textit{i.e.,} $\beta_g = K$) described in Section~\ref{subsec:prior-application}.
Here, we relax this fixed assumption and instead adopt the following hierarchical prior, which allows $\beta_g$ to vary across groups with a mean of $K$:
\begin{gather*}
\mu_{\beta} = K, \ \ \sigma_\beta \sim N^{+}(0, 1000^2).
\end{gather*}

\begin{figure*}[!t]
\centering
\subfloat[Threshold at 30,000]{\includegraphics[width=0.50\textwidth]{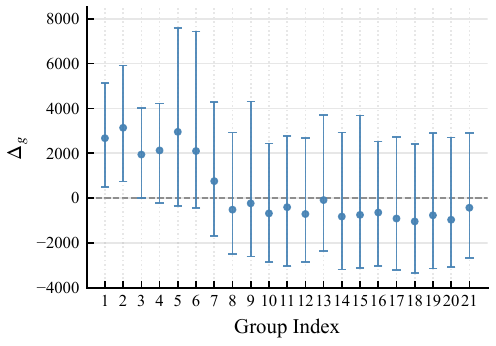} \label{fig:th_30000_appendix}}
\hfill
\subfloat[Threshold at 50,000]{\includegraphics[width=0.50\textwidth]{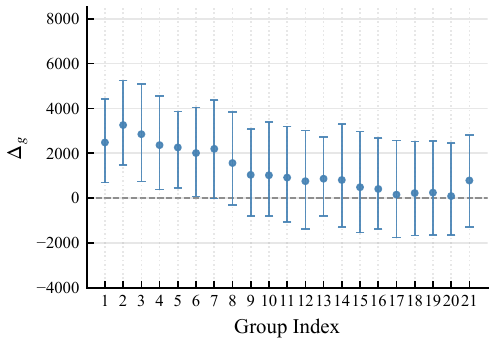} \label{fig:th_50000_appendix}}
\bigskip 
\centering
\subfloat[Threshold at 70,000]{\includegraphics[width=0.50\textwidth]{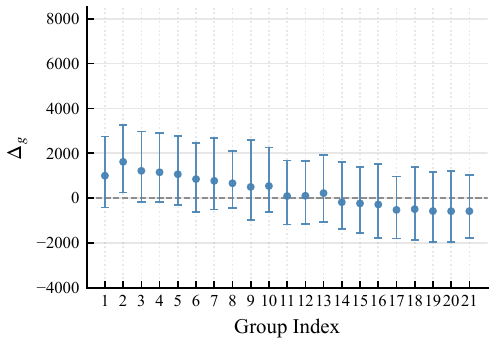} \label{fig:th_70000_appendix}}
\caption{Point estimates and 90\% credible intervals of $\Delta_g$ obtained by the proposed method HBMTM at each threshold under the alternative prior specification.}
\label{fig:sensitivity-analysis-results}
\end{figure*}

Figure~\ref{fig:sensitivity-analysis-results} provides the estimated causal effects $\Delta_g$ for each subgroup under this alternative specification.
As shown in Figure~\ref{fig:sensitivity-analysis-results}, our sensitivity analyses confirm this primary finding. 
First, a clear positive causal effect was consistently observed in all model specifications for subgroups with prior-month spending near or slightly above the threshold. For some of these groups, the 90\% HDI remained entirely in the positive range, suggesting that the effect is statistically robust.
Second, the trend of diminishing effects for high-spending subgroups also proved robust, though this pattern was prominent at the K1 and K3 thresholds and less apparent at the K2 threshold. 
Therefore, the qualitative conclusions from our main analysis hold under this alternative prior specification.

\end{document}